\documentstyle[12pt]{article}
\input epsf

\begin{document}

\begin{titlepage}

\begin{flushright}
Freiburg--THEP 97--04\\
UM--TH--97--06\\
February 1997
\end{flushright}
\vspace{2.0cm}

\begin{center}
\large\bf
{\LARGE\bf Massive two--loop integrals in renormalizable theories}\\[2cm]
\rm
{Adrian Ghinculov$^a$ and York--Peng Yao$^b$}\\[.5cm]

{\em $^a$Albert--Ludwigs--Universit\"{a}t Freiburg,
                               Fakult\"{a}t f\"{u}r Physik,}\\
      {\em Hermann--Herder Str.3, D-79104 Freiburg, Germany}\\[.2cm]
{\em $^b$Randall Laboratory of Physics, University of Michigan,}\\
      {\em Ann Arbor, Michigan 48109--1120, USA}\\[3.0cm]
      
\end{center}
\normalsize

\begin{abstract}

We propose a framework for calculating 
two--loop Feynman diagrams which appear within 
a renormalizable theory in the general mass case 
and at finite external momenta. Our approach is a combination of 
analytical results and of high accuracy numerical integration,
similar to a method proposed previously \cite{2loop:2} 
for treating diagrams without numerators. We reduce all possible 
tensor structures to a small set of scalar integrals, 
for which we provide integral representations in terms of four basic 
functions. The algebraic part is suitable for implementing in a 
computer program for the automatic generation and evaluation of Feynman 
graphs. The numerical part is essentially the same as in the case of 
Feynman diagrams without numerators.
\end{abstract}

\vspace{3cm}

\end{titlepage}


\title{Massive two--loop integrals in renormalizable theories}

\author{Adrian Ghinculov$^a$ and York--Peng Yao$^b$}

\date{{\em $^a$Albert--Ludwigs--Universit\"{a}t Freiburg,
                               Fakult\"{a}t f\"{u}r Physik,}\\
      {\em Hermann--Herder Str.3, D-79104 Freiburg, Germany}\\[.2cm]
      {\em $^b$Randall Laboratory of Physics, University of Michigan,}\\
      {\em Ann Arbor, Michigan 48109--1120, USA}}

\maketitle

\begin{abstract}

We propose a framework for calculating 
two--loop Feynman diagrams which appear within 
a renormalizable theory in the general mass case 
and at finite external momenta. Our approach is a combination of 
analytical results and of high accuracy numerical integration,
similar to a method proposed previously \cite{2loop:2} 
for treating diagrams without numerators. We reduce all possible 
tensor structures to a small set of scalar integrals, 
for which we provide integral representations in terms of four basic 
functions. The algebraic part is suitable for implementing in a 
computer program for the automatic generation and evaluation of Feynman 
graphs. The numerical part is essentially the same as in the case of 
Feynman diagrams without numerators.
\end{abstract}


\section{Introduction}

The importance of developing techniques to calculate 
radiative corrections can hardly be overemphasized.
The measurements at existing high energy colliders, 
like LEP and the SLC, are clearly sensitive to one--loop 
electroweak effects. As their precision increases, 
analyses at two--loop level may become necessary \cite{preccalc}.

The problem of calculating one--loop Feynman diagrams was in 
principle solved completely long time ago by 't Hooft and Veltman 
\cite{veltthooft}, and by Passarino and Veltman \cite{passvelt}. 
For a review, see for instance ref. \cite{denner}, and 
refs. \cite{kniehlrev,hollik} for some applications. 
As it is well--known, all one--loop diagrams are expressible 
in terms of Spence functions. In the general mass case, 
however, the numerical evaluation of the resulting expressions 
may be tricky \cite{vermold}. This is because one needs to take 
care that the functions involved remain on the correct Riemann sheet,
and because one has to control potentially large numerical cancellations.

No similar general solution was available so far at two--loop level
in the general massive case, although a lot of effort
was devoted to solving two--loop massive
integrals at finite external momenta.  There are two issues involved, given
a general diagram with particles with spins: First, the numerator of the
amplitude
usually has non-trivial tensorial structures.  It is convenient
for us to decompose the amplitude into various tensorial constructs made of
external momenta and metric tensors, multiplied by scalar functions, which
in turn must be related back to the original Feynman integral by certain
projections. Second, one should advance an algorithm by which one can
evaluate these generalized scalar integrals, which have non-trivial
numerators, in a systematic and, better yet, universal way. 

Regarding the evaluation of two--loop
scalar integrals, a lot of work was done in this direction lately,
and it has become clear that in general
the integrals involved at two--loop level are not expressible analytically
in terms of well--known and easy to evaluate functions like polylogarithms,
as is the case with one--loop diagrams.  Partly numerical approaches
seem unavoidable in the general mass case.  Only in some rare situations is
an analytical solution in terms of special functions known. Such an example is ref.
\cite{lauricella}, where a certain two--loop self--energy diagram was related
to the Lauricella functions. A large number of approaches were proposed
which work for specific topologies or mass combinations - see, for instance,
refs. \cite{lauricella}---\cite{fujimoto} and references therein. In most
cases, these works deal with Feynman integrals with simplified couplings
and numerators. Some of these approaches can yield precise numerical results for 
the diagrams to which they apply; by combining several such methods, one was
able to apply them to actual physical processes. While they do hold promise to be
extendable to cover more complicated scalar 
integrals which may appear in processes
involving particles with spin, derivative couplings, 
and more external momenta, it is fair to say that this program 
has not been completed so far.  Moreover, 
for calculating a physical process, such approaches would involve 
separate methods to treat individual graphs. This implies a considerable
amount of work which grows very fast with the complexity of the process
considered.

Fewer results exist concerning
the reduction of two--loop Feynman graphs
to scalar integrals. In ref. \cite{weiglein:1}
it was shown that the two--loop self--energy
diagrams can be reduced to scalar integrals without numerators. This result
applies only to two--point functions. It was used in conjunction
with numerical integration in ref.
\cite{weiglein:2} for
calculating certain two--loop contributions to the $\rho$ parameter.
The approach of ref. \cite{weiglein:1} only refers to propagator diagrams. Ref.
\cite{tarasov}, which also deals with propagator
diagrams, might be extendable to some more complicated diagrams, like certain
3-point functions in simplified kinematic cases, and the recurrence relations
of ref. \cite{tarasov} perhaps can be solved numerically in more complicated cases,
as suggested in the Conclusions of ref. \cite{tarasov}.

Compared to the approaches briefly discussed above, in this paper 
we present a general universal framework which applies
to any n-particle 2-loop diagram with arbitrary renormalizable
dynamics and any combination of masses and external momenta. Strictly speaking,
our formulae can be used directly for calculating one given diagram only if that
diagram is free of mass singularities. However, in actual calculations this 
can be often circumvented by isolating the singularity analytically at the level of 
the integrand, by considering infrared finite sums of diagrams, or by introducing 
a mass regulator, as was possible for instance in ref. \cite{2loop:3}.  
In our approach, all diagrams are treated using the same algorithm, and for this reason
the whole calculation, including the generation of the relevant diagrams,
can be authomatized. This was done in ref. \cite{2loop:2} in a simplified case
of $H \rightarrow WW$ decay, which involves 
no tensor structures in the numerator,
and is a subset of what we shall discuss in this paper. This can be compared
with ref. \cite{frink}, whose authors used separate methods for individual diagrams
to perform the same calculation, and which is so far the state--of--the--art
of physical calculations by using these methods.

Within the framework which we propose, we shall show that in any renormalizable 
theory all two--loop diagrams can be expressed in terms of a few
scalar functions. To our best knowledge, these functions 
cannot be expressed analytically in terms of known functions, in a way which 
would facilitate their numerical evaluation. There should be in principle a 
connection with generalized hypergeometric functions, but it's not clear that 
this would lead to an efficient evaluation of these functions. Instead of 
looking for an analytical result, we derive one--dimensional integral 
representations for these scalar functions. They can be expressed in terms
of four simple functions. The structure of these integral representations 
is a generalization of the results derived in ref. \cite{2loop:2} 
for the case of 
two--loop diagrams without numerators. Such integral representations can be 
used for calculating  numerically the necessary functions very efficiently. 
This was shown in refs. \cite{2loop:2,2loop:Htott} for two--point functions 
and in ref. \cite{2loop:3} for three--point functions.

Compared to the case of diagrams without numerators, two new
but similar structures may appear in a renormalizable theory, like the 
standard model. We note that in nonrenormalizable theories more structures 
are allowed. As it will become clear from our discussion, these 
additional structures can be treated along the same lines, and 
will result in similar functions.

{\bf Added Note} 
After the submission of this article, a related work by O.V. Tarasov appeared
\cite{tarasov}. This
work makes use of some special properties (dimensional
shifts), due to Schwinger representation of propagators and was able to
express any two-loop {\em propagator} diagrams by a minimal set
of four master integrals.  It is further asserteded that this result does
not depend on the renormalizability of the theory. On the other hand,
our assertion is that {\em any two-loop n-point diagram} can be
expressed by four basic functions, as long as the theory is renormalizable.
We have not specifically looked into propagator diagrams to see if there
are extra symmetries or recurrence relations such that this result still
holds for nonrenormalizable theories. Furthermore, we were not primarily
interested in identifying a minimal set of basic functions, but rather
in obtaining a framework which leads to expressions convenient to evaluate numerically.
Thus, our remark that there is no finite set of basis integrals
for non-renormalizable theories should be taken only as a most direct
inference from our work so far.


\section{Reduction to scalar invariants}

In this section we show that the calculation of two--loop Feynman diagrams
with any tensor structure at the numerator can be reduced to the 
evaluation of a class of scalar invariants.

Any two--loop diagram without numerators can be written as an integral 
over scalar integrals of the type \cite{2loop:2}:

\begin{eqnarray}
\lefteqn{G(m_{1},\alpha_{1};
           m_{2},\alpha_{2};
           m_{3},\alpha_{3};k^2)  \, =}  \nonumber \\ 
& &    \;\;\;\;\;\;\;\;\;
       \int d^{n}p\,d^{n}q\, 
       \frac{1}{
             (p^{2}+m_{1}^{2})^{\alpha_{1}} \,
             (q^{2}+m_{2}^{2})^{\alpha_{2}} \,
             [(r+k)^{2}+m_{3}^{2}]^{\alpha_{3}}
	    }
    \; \; ,
\end{eqnarray}
where all momenta are Euclidian.

This form is obtained after combining all propagators which contain the 
same loop momentum $p$, $q$ or $r \equiv p+q$ 
by introducing Feynman parameters.
At their turn, all $G$ functions can be derived from two basic functions
${\cal F}$ and ${\cal G}$. For a discussion of the properties of these
functions see refs. \cite{2loop:2,2loop:3}.
To calculate a Feynman diagram, one typically has to perform a numerical
integration of. eq. 1. The maximum dimension of this integration is the 
number of propagators of the Feynman diagram minus three.

A Feynman diagram may contain non--trivial tensor structures at 
the numerator. For all these cases, tensor structures in the original
Feynman diagram will result in tensor integrals of the type:

\begin{equation}
   \int d^{n}p\,d^{n}q\, 
       \frac{p^{\mu_1} \ldots p^{\mu_i} q^{\mu_{i+1}} \ldots q^{\mu_j}}{
             (p^{2}+m_{1}^{2})^{\alpha_{1}} \,
             (q^{2}+m_{2}^{2})^{\alpha_{2}} \,
             [(r+k)^{2}+m_{3}^{2}]^{\alpha_{3}}
	    }
    \; \; .
\end{equation}

The calculation of such tensor integrals can be reduced to the 
calculation of scalar integrals of the following form:

\begin{equation}
    \int d^{n}p\,d^{n}q\, 
       \frac{(p \cdot k)^a (q \cdot k)^b}{
             (p^{2}+m_{1}^{2})^{\alpha_{1}} \,
             (q^{2}+m_{2}^{2})^{\alpha_{2}} \,
             [(r+k)^{2}+m_{3}^{2}]^{\alpha_{3}}
	    }
    \; \; .
\end{equation}

To show this, let us introduce the transverse components of the 
loop momenta $p$ and $q$:

\begin{eqnarray}
 p_{\perp}^{\mu} & = &  p^{\mu} - \frac{p \cdot k}{k^2} k^{\mu} 
 \nonumber \\
 q_{\perp}^{\mu} & = &  q^{\mu} - \frac{q \cdot k}{k^2} k^{\mu} 
    \; \; .
\end{eqnarray}

With these notations, eq. 2 results in a sum of terms with 
the following structure:

\begin{equation}
I^{\mu_1 \cdots \mu_{j^{\prime}}}
              =
    \int d^{n}p\,d^{n}q\, 
       \frac{p^{\mu_1}_{\perp} \ldots p^{\mu_{i^{\prime}}}_{\perp}  
         q^{\mu_{i^{\prime}+1}}_{\perp} \ldots q^{\mu_{j^{\prime}}}_{\perp}
             P(p \cdot k , q \cdot k)}{
             (p^{2}+m_{1}^{2})^{\alpha_{1}} \,
             (q^{2}+m_{2}^{2})^{\alpha_{2}} \,
             [(r+k)^{2}+m_{3}^{2}]^{\alpha_{3}}
	    }
    \; \; ,
\end{equation}
where $P$ is some poynomial function of $p \cdot k$ and $q \cdot k$.
One can convince oneself that the $1/k^2$ singularities introduced
in these expressions by the decomposition of eqns. 4 are superfluous.
They disappear completely when one adds the transverse and longitudinal
parts together. We shall show this explicitly in future publications
devoted to calculations of specific processes.

Integrals of this type vanish for odd $j^{\prime}$. This is because
the tensor $I^{\mu_1 \cdots \mu_{j^{\prime}}}$ is transverse with respect 
to all its indices. At the same time, the only vector available 
for constructing $I^{\mu_1 \cdots \mu_{j^{\prime}}}$ after the $p$
and $q$ integrations in eq. 4 are carried out 
is the external momentum $k^{\mu}$.
If $j^{\prime}=2n+1$, after using $k$ and the metric tensor 
$g^{\mu \nu}$ to support $2n$
transverse indices, there is one free index left, 
which must be carried by $k$. However, this last index 
cannot be made transverse, and thus our assertion follows.


As an example, one has the following relation:

\begin{eqnarray}
\lefteqn{\int d^{n}p\,d^{n}q\, 
       \frac{p^{\mu} P(p \cdot k , q \cdot k)}{
             (p^{2}+m_{1}^{2})^{\alpha_{1}} \,
             (q^{2}+m_{2}^{2})^{\alpha_{2}} \,
             [(r+k)^{2}+m_{3}^{2}]^{\alpha_{3}}
	    }  
    \, =}  \nonumber \\ 
& &  \;\;\;\;\;\;\;\;\;\;\;\;
     \frac{k^{\mu}}{k^2}
\int d^{n}p\,d^{n}q\, 
       \frac{p \cdot k P(p \cdot k , q \cdot k)}{
             (p^{2}+m_{1}^{2})^{\alpha_{1}} \,
             (q^{2}+m_{2}^{2})^{\alpha_{2}} \,
             [(r+k)^{2}+m_{3}^{2}]^{\alpha_{3}}
	    }
    \; \; .
\end{eqnarray}
This is because the following relation holds:

\begin{equation}
\int d^{n}p\,d^{n}q\, 
       \frac{p_{\perp}^{\mu} P(p \cdot k , q \cdot k)}{
             (p^{2}+m_{1}^{2})^{\alpha_{1}} \,
             (q^{2}+m_{2}^{2})^{\alpha_{2}} \,
             [(r+k)^{2}+m_{3}^{2}]^{\alpha_{3}}
	    }
 = 0
    \; \; .
\end{equation}

If $j^{\prime}$ in eq. 5 is even, one can always decompose 
$I^{\mu_1 \cdots \mu_{j^{\prime}}}$ into scalar invariants of 
the type in eq. 3 by using the transversality of 
$I^{\mu_1 \cdots \mu_{j^{\prime}}}$. 

For instance, apart from a redefinition of the loop momenta
$p$ and $q$, two combinations are possible for $j^{\prime} = 2$:

\begin{eqnarray}
\int d^{n}p\,d^{n}q\, 
       \frac{p^{\mu_1}_{\perp} p^{\mu_2}_{\perp}  P(p \cdot k , q \cdot k)}{
             (p^{2}+m_{1}^{2})^{\alpha_{1}} \,
             (q^{2}+m_{2}^{2})^{\alpha_{2}} \,
             [(r+k)^{2}+m_{3}^{2}]^{\alpha_{3}}
	    } & = & ( g^{\mu_1 \mu_2} - \frac{k^{\mu_1} k^{\mu_2}}{k^2} ) A 
\nonumber \\ 
\int d^{n}p\,d^{n}q\, 
       \frac{p^{\mu_1}_{\perp} q^{\mu_2}_{\perp}  P(p \cdot k , q \cdot k)}{
             (p^{2}+m_{1}^{2})^{\alpha_{1}} \,
             (q^{2}+m_{2}^{2})^{\alpha_{2}} \,
             [(r+k)^{2}+m_{3}^{2}]^{\alpha_{3}}
	    } & = & ( g^{\mu_1 \mu_2} - \frac{k^{\mu_1} k^{\mu_2}}{k^2} ) B 
    \; \; ,
\end{eqnarray}
where

\begin{eqnarray}
A & = &  \frac{1}{n-1}
\int d^{n}p\,d^{n}q\, 
       \frac{ p_{\perp}^2  
             P(p \cdot k , q \cdot k)}{
             (p^{2}+m_{1}^{2})^{\alpha_{1}} \,
             (q^{2}+m_{2}^{2})^{\alpha_{2}} \,
             [(r+k)^{2}+m_{3}^{2}]^{\alpha_{3}}}
\nonumber \\ 
B & = &  \frac{1}{n-1}
\int d^{n}p\,d^{n}q\, 
       \frac{ (p_{\perp} \cdot q_{\perp}) 
             P(p \cdot k , q \cdot k)}{
             (p^{2}+m_{1}^{2})^{\alpha_{1}} \,
             (q^{2}+m_{2}^{2})^{\alpha_{2}} \,
             [(r+k)^{2}+m_{3}^{2}]^{\alpha_{3}}}
    \; \; .
\end{eqnarray}

The scalar invariants $A$ and $B$ can be further reduced 
to functions of the form of eq. 3 by using the relations:

\begin{eqnarray}
p_{\perp}^2               & = & p^2       - \frac{1}{k^2} \, 
                                                 (p \cdot k)^2
\nonumber \\ 
p_{\perp} \cdot q_{\perp} & = & p \cdot q - \frac{1}{k^2} \,
                                      (p \cdot k)(q \cdot k)
\nonumber \\ 
p \cdot q                 & = & \frac{1}{2} \, (r^2 - p^2 - q^2)
\nonumber 
\end{eqnarray}
and by partial fractioning the resulting expressions.

The corresponding relations for  $j^{\prime} = 4$ are given in Appendix A.

Obviously, this procedure can be extended to higher order
tensor structures which may occur in more complicated calculations. 
At this point, any two--loop Feynman diagram can be expressed
in terms of scalar integrals of the type of eq. 3.


\section{Recursion relations}

We have shown in the previous section that the problem of calculating any 
two--loop Feynman diagram reduces to treating scalar integrals of the type:

\begin{eqnarray}
\lefteqn{
  {\cal P}^{a \, b}_{\alpha_1 \, \alpha_2 \, \alpha_3} (m_1,m_2,m_3;k^2)
    \, \equiv}  \nonumber \\ 
& & \;\;\;\;\;\;\;\;\;\;\;\;\;\;\;\;\;\;\;\;\;\;\;\;
       \int d^{n}p\,d^{n}q\, 
       \frac{(p \cdot k)^a (q \cdot k)^b}{
             (p^{2}+m_{1}^{2})^{\alpha_{1}} \,
             (q^{2}+m_{2}^{2})^{\alpha_{2}} \,
             [(r+k)^{2}+m_{3}^{2}]^{\alpha_{3}}
	    }
    \; \; ,
\end{eqnarray}
where $\alpha_1,\alpha_2,\alpha_3 \geq 1$ and $a,b \geq 0$.

Not all ${\cal P}^{a \, b}_{\alpha_1 \, \alpha_2 \, \alpha_3}$
are independent. There are recursion relations
between functions with different indices. 
In this section we show that all
${\cal P}^{a \, b}_{\alpha_1 \, \alpha_2 \, \alpha_3}$ functions
which can appear in a two--loop calculation within a renormalizable
theory can be derived by differentiation from a limited number 
of basic functions.

As a first step, one notices that one can increase the indices
$\alpha_1$, $\alpha_2$ and $\alpha_3$ of
${\cal P}^{a \, b}_{\alpha_1 \, \alpha_2 \, \alpha_3}$
by one unit by differentiating with respect to the mass arguments:

\begin{equation}
{\cal P}^{a \, b}_{\alpha_1+1 \, \alpha_2 \, \alpha_3}(m_1,m_2,m_3;k^2) =
 - \frac{1}{\alpha_1} \frac{\partial}{\partial m_1^2}
{\cal P}^{a \, b}_{\alpha_1   \, \alpha_2 \, \alpha_3}(m_1,m_2,m_3;k^2)
    \; \; ,
\end{equation}
and similarly for $\alpha_2$ and $\alpha_3$. 

Next, by differentiation with respect to the external momentum
$k^{\mu}$, one obtains the following relations:

\begin{eqnarray}
  {\cal P}^{a+1 \, b}_{\alpha_1+1 \, \alpha_2 \, \alpha_3} & = &
  \frac{1}{2 \; \alpha_1} \left[ 2 k^2 \frac{\partial}{\partial k^2} 
                                 - (a+b)
                          \right]
  {\cal P}^{a \, b}_{\alpha_1 \, \alpha_2 \, \alpha_3}
  + \frac{a k^2}{2 \alpha_1} 
  {\cal P}^{a-1 \, b}_{\alpha_1 \, \alpha_2 \, \alpha_3}
\nonumber \\
  {\cal P}^{a \, b+1}_{\alpha_1 \, \alpha_2+1 \, \alpha_3} & = &
  \frac{1}{2 \; \alpha_2} \left[ 2 k^2 \frac{\partial}{\partial k^2} 
                                 - (a+b)
                          \right]
  {\cal P}^{a \, b}_{\alpha_1 \, \alpha_2 \, \alpha_3}
  + \frac{b k^2}{2 \alpha_2} 
  {\cal P}^{a \, b-1}_{\alpha_1 \, \alpha_2 \, \alpha_3}
    \; \; ,
\end{eqnarray}
and

\begin{equation}
   \left[ 2 k^2 \frac{\partial}{\partial k^2} - (a+b)
  \right]
  {\cal P}^{a \, b}_{\alpha_1 \, \alpha_2 \, \alpha_3}  = 
  - 2 \alpha_3
   \left[ k^2
   {\cal P}^{a   \, b  }_{\alpha_1 \, \alpha_2 \, \alpha_3+1} +
   {\cal P}^{a+1 \, b  }_{\alpha_1 \, \alpha_2 \, \alpha_3+1} +
   {\cal P}^{a   \, b+1}_{\alpha_1 \, \alpha_2 \, \alpha_3+1}
  \right]
    \; \; .
\end{equation}

In eqns. 12, the functions 
${\cal P}^{a=-1 \, b}_{\alpha_1   \, \alpha_2   \, \alpha_3}$
or
${\cal P}^{a \, b=-1}_{\alpha_1   \, \alpha_2   \, \alpha_3}$,
which may appear if either $a$ or $b$ vanish, are defined to be zero.
These relations allow one to increase the upper and the lower indices
by one unit simultaneously.

In the following we will call $\alpha_1+\alpha_2+\alpha_3-a-b$
the "degree" of the function 
${\cal P}^{a \, b}_{\alpha_1 \, \alpha_2 \, \alpha_3}$.
Eqns. 11 can be used for increasing the degree of 
${\cal P}^{a \, b}_{\alpha_1 \, \alpha_2 \, \alpha_3}$.
Eqns. 12 essentially relate functions of the same degree, 
while increasing simultaneously the upper and the lower indices.

Hence, it is important to look at the functions of lowest possible 
degree in order to identify a limited set of functions from which one
can derive all the other.

The allowed range of the degree of the functions
${\cal P}^{a \, b}_{\alpha_1 \, \alpha_2 \, \alpha_3}$
which can appear in two--loop amplitudes depends 
on the theory under consideration. For instance, in the case
of the linear sigma model the minimum degree is three 
\cite{2loop:2,2loop:3}. 
There is no upper limit, since the degree can be 
increased indefinitely by adding external legs to the diagram.

In renormalizable theories, the degree of two--loop functions
has always a lower bound. To see this, one notes that powers 
of the loop momenta appear in the numerator of 
a Feynman diagram only from the fermion
propagators and from derivative couplings. If one demands the
theory to be renormalizable, one can only have fermion--boson 
couplings with no derivative, and trilinear boson couplings
which have at most one derivative. One can convince oneself
that the minimum degree of a two--loop function is one, 
and is attained by the vacuum diagram with two fermion 
propagators and one boson propagator 
shown in fig. 1. By adding external bosonic 
lines or pairs of fermionic
lines to a diagram one can only increase the degree of the diagram.

\begin{figure}
\hspace{4.5cm}
    \epsfxsize = 5cm
    \epsffile{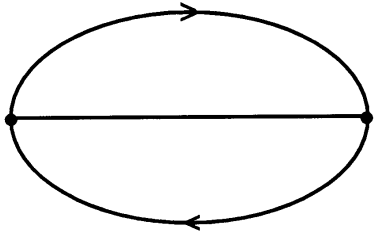}
\caption{Two--loop vacuum diagram with one boson and two fermion 
         propagators, which leads to a ${\cal P}$ function
         of minimum degree (one). Vacuum diagrams are expressed
	 analytically in terms of Spence functions, but one can
	 obtain nontrivial ${\cal P}$ functions of degree one
	 by adding external bosonic lines to this diagram.}
\end{figure}

Therefore, by using eqns. 11 and 12 one can derive all functions
${\cal P}^{a \, b}_{\alpha_1 \, \alpha_2 \, \alpha_3}$
which may appear in renormalizable theories from those
with $\alpha_1 = \alpha_2 = \alpha_3 = 1$ and $a+b = 0, 1, 2$.
Nevertheless, we choose to use the following set instead:

\begin{equation}
\begin{array}{lllll} 
\mbox{degree} = 4 :                                   \;\;\;\;
                    & {\cal P}^{0 \, 0}_{2 \, 1 \, 1} \;\;\;\;
                    &  
                    &  
                    &   
\\
\mbox{degree} = 3 :                                   \;\;\;\;
                    & {\cal P}^{1 \, 0}_{2 \, 1 \, 1} \;\;\;\;
                    & {\cal P}^{0 \, 1}_{2 \, 1 \, 1} \;\;\;\;
                    &  
                    &   
\\
\mbox{degree} = 2 :                                   \;\;\;\;
                    & {\cal P}^{2 \, 0}_{2 \, 1 \, 1} \;\;\;\;
                    & {\cal P}^{1 \, 1}_{2 \, 1 \, 1} \;\;\;\;
                    & {\cal P}^{0 \, 2}_{2 \, 1 \, 1} \;\;\;\;
                    &   
\\
\mbox{degree} = 1 :                                   \;\;\;\;
                    & {\cal P}^{3 \, 0}_{2 \, 1 \, 1} \;\;\;\;
                    & {\cal P}^{2 \, 1}_{2 \, 1 \, 1} \;\;\;\;
                    & {\cal P}^{1 \, 2}_{2 \, 1 \, 1} \;\;\;\;
                    & {\cal P}^{0 \, 3}_{2 \, 1 \, 1} \;\;\;\;
\end{array}
\end{equation}
The reason is that for these functions one can obtain simpler 
integral representations than for the functions with 
$\alpha_1 = \alpha_2 = \alpha_3 = 1$. 
The functions with $\alpha_1 = \alpha_2 = \alpha_3 = 1$
are then derived from the set of eq. 14 by using the "partial $p$" 
operation of 't Hooft and Veltman \cite{veltthooft}.

By applying partial $p$ to 
${\cal P}^{a \, b}_{\alpha_1 \, \alpha_2 \, \alpha_3}$, one finds
the following relation:

\begin{eqnarray}
  {\cal P}^{a \, b}_{\alpha_1 \, \alpha_2 \, \alpha_3} 
    & = &
  - \frac{1}{n-(\alpha_1 + \alpha_2 + \alpha_3) + (a+b)/2 }
  \left\{
 \alpha_1 m_1^2  {\cal P}^{a \, b}_{\alpha_1+1 \, \alpha_2 \, \alpha_3} +
 \alpha_2 m_2^2  {\cal P}^{a \, b}_{\alpha_1 \, \alpha_2+1 \, \alpha_3} 
  \right.
\nonumber \\
    &   &
  \left.  \;\;\;\;\;\;\;\;\;\;\;\;\;\;\;\;\;
 + \alpha_3 m_3^2  {\cal P}^{a \, b}_{\alpha_1 \, \alpha_2 \, \alpha_3+1} -
   \frac{1}{2}
   \left[ 2 k^2 \frac{\partial}{\partial k^2} - (a+b)
  \right]
       {\cal P}^{a \, b}_{\alpha_1 \, \alpha_2 \, \alpha_3}
 \right\}
    \; \; ,
\end{eqnarray}
and this can be used together with eqns. 12 or 13 
for expressing the functions with
$\alpha_1 = \alpha_2 = \alpha_3 = 1$ in terms of the set 
in eq. 14 when they are needed.

Note that not all functions in eq. 14 are independent. Some
of them are related by loop momentum redefinitions. For instance,
the following relation holds:

\begin{eqnarray}
\lefteqn{
{\cal P}^{1 \, 0}_{2 \, 1 \, 1} (m_1,m_2,m_3;k^2)
+
{\cal P}^{0 \, 1}_{2 \, 1 \, 1} (m_1,m_2,m_3;k^2)
    \, =}  \nonumber \\ 
& & \;\;\;\;\;\;\;\;\;\;\;\;\;\;\;\;\;\;\;\;\;\;\;\;
- \left[
{\cal P}^{0 \, 1}_{2 \, 1 \, 1} (m_1,m_3,m_2;k^2)
+ k^2
{\cal P}^{0 \, 0}_{2 \, 1 \, 1} (m_1,m_3,m_2;k^2)
 \right]
    \; \; .
\end{eqnarray}

At the same time, some of these functions may actually never 
appear directly from a Feynman diagram, as is the case with 
${\cal P}^{0 \, 3}_{2 \, 1 \, 1}$. However, they may appear
indirectly when using eqns. 12, 13 or 15. For this reason we prefer
not to discard them at this point.

The set of functions in eq. 14 is an extension of the functions
${\cal F}$ and ${\cal G}$ of ref. \cite{2loop:2}. 
${\cal F}$ and ${\cal G}$
are enough for treating all diagrams without numerators. This is the
case for instance with the linear sigma model. As already mentioned,
the lowest degree which may appear in the linear sigma model is three.
Therefore, only the first two lines of eq. 14 are involved. In
fact, ${\cal P}^{0 \, 0}_{2 \, 1 \, 1} = {\cal G}$, and
${\cal P}^{1 \, 0}_{2 \, 1 \, 1} = -{\cal F}$. The remaining
degree three function, ${\cal P}^{0 \, 1}_{2 \, 1 \, 1}$,
never appears directly in this case from a Feynman diagram.

One also notices that eq. 15 is a generalization of eq. 11
of ref. \cite{2loop:2}.

Finally, a remark on nonrenormalizable theories is in order. One
can construct two--loop Feynman diagrams of degree lower than one
if one allows for nonrenormalizable interactions. As an example,
the two--loop sunset self--energy in a four--fermion interaction theory
is of degree zero. In nonrenormalizable theories it is not always
possible to restrict the degree of the 
${\cal P}^{a \, b}_{\alpha_1 \, \alpha_2 \, \alpha_3}$
functions and thus to identify a finite set of basic functions
which generate all the other. In the following section we will
derive one--dimensional integral representations for the functions
of eq. 14. As it will become clear, it is possible to derive similar
integral representations for functions with $a+b > 3$ as well.
Thus, specific calculations in nonrenormalizable theories can be performed,
too. One only needs to introduce additional functions along with
those of eq. 14.


\section{Integral representations}

We have seen in the previous sections that any two--loop Feynman 
diagram in a renormalizable theory can be expressed in terms of 
the ten functions of eq. 14, not all of them independent.

As said, our aim here is to express the ultraviolet finite parts of 
a calculation in a general kinematical situation in a form
which would facilitate their numerical evaluation.
The reader may know that, already at the level of two--point functions, 
the self--energy sunset diagram is known to be related to the 
Lauricella functions \cite{lauricella}. These functions are not 
straightforward to evaluate numerically. The best way to do that 
appears to be by means of integral representations \cite{bauberger}. 
For more complicated two--loop diagrams no analytical results 
are known so far. Besides, we want to deal with more realistic
situations, where non--trivial numerators are present in the integrands.

Hence, in this section we derive one--dimensional integral 
representations of the ten functions in eq. 14. Such integral 
representations can be used for a fast and accurate numerical 
evaluation of these functions.  

It is convenient to replace the set of ten functions
${\cal P}^{a \, b}_{2 \, 1 \, 1}(m_1,m_2,m_3;k^2)$ 
of eq. 14 by the following equivalent set 
${\cal H}_i (m_1,m_2,m_3;k^2)$:

\begin{eqnarray}
   {\cal P}^{0 \, 0}_{2 \, 1 \, 1} \; \rightarrow {\cal H}_1 
   & = &
    \int d^{n}p\,d^{n}q\, 
    \frac{1}{
          [(p+k)^2+m_{1}^{2}]^2 \,
          (q^2    +m_{2}^{2}) \,
          (r^2    +m_{3}^{2})         }   
  \nonumber \\
   {\cal P}^{1 \, 0}_{2 \, 1 \, 1} \; \rightarrow {\cal H}_2
   & = &
    \int d^{n}p\,d^{n}q\, 
    \frac{ p \cdot k }{
          [(p+k)^2+m_{1}^{2}]^2 \,
          (q^2    +m_{2}^{2}) \,
          (r^2    +m_{3}^{2})         }   
  \nonumber \\
   {\cal P}^{0 \, 1}_{2 \, 1 \, 1} \; \rightarrow {\cal H}_3
   & = &
    \int d^{n}p\,d^{n}q\, 
    \frac{ q \cdot k }{
          [(p+k)^2+m_{1}^{2}]^2 \,
          (q^2    +m_{2}^{2}) \,
          (r^2    +m_{3}^{2})         }   
  \nonumber \\
   {\cal P}^{2 \, 0}_{2 \, 1 \, 1} \; \rightarrow {\cal H}_4
   & = &
    \int d^{n}p\,d^{n}q\, 
    \frac{ (p \cdot k)^2 
         - \frac{1}{n} k^2 p^2 }{
          [(p+k)^2+m_{1}^{2}]^2 \,
          (q^2    +m_{2}^{2}) \,
          (r^2    +m_{3}^{2})         }   
  \nonumber \\
   {\cal P}^{1 \, 1}_{2 \, 1 \, 1} \; \rightarrow {\cal H}_5
   & = &
    \int d^{n}p\,d^{n}q\, 
    \frac{ (p \cdot k) (q \cdot k) 
         - \frac{1}{n} k^2 (q \cdot p)  }{
          [(p+k)^2+m_{1}^{2}]^2 \,
          (q^2    +m_{2}^{2}) \,
          (r^2    +m_{3}^{2})         }   
  \nonumber \\
   {\cal P}^{0 \, 2}_{2 \, 1 \, 1} \; \rightarrow {\cal H}_6
   & = &
    \int d^{n}p\,d^{n}q\, 
    \frac{ (q \cdot k)^2 
         - \frac{1}{n} k^2 q^2  }{
          [(p+k)^2+m_{1}^{2}]^2 \,
          (q^2    +m_{2}^{2}) \,
          (r^2    +m_{3}^{2})         }   
  \nonumber \\
   {\cal P}^{3 \, 0}_{2 \, 1 \, 1} \; \rightarrow {\cal H}_7
   & = &
    \int d^{n}p\,d^{n}q\, 
    \frac{ (p \cdot k)^3 
         - \frac{3}{n+2} k^2 p^2 (p \cdot k)  }{
          [(p+k)^2+m_{1}^{2}]^2 \,
          (q^2    +m_{2}^{2}) \,
          (r^2    +m_{3}^{2})         }   
  \nonumber \\
   {\cal P}^{2 \, 1}_{2 \, 1 \, 1} \; \rightarrow {\cal H}_8
   & = &
    \int d^{n}p\,d^{n}q\, 
    \frac{ (p \cdot k)^2 (q \cdot k) 
         -   \frac{3}{n+2} k^2 p^2 (q \cdot k)   }{
          [(p+k)^2+m_{1}^{2}]^2 \,
          (q^2    +m_{2}^{2}) \,
          (r^2    +m_{3}^{2})         }   
  \nonumber \\
   {\cal P}^{1 \, 2}_{2 \, 1 \, 1} \; \rightarrow {\cal H}_9 
   & = &
    \int d^{n}p\,d^{n}q\, 
    \frac{ (p \cdot k) (q \cdot k)^2 
        -  \frac{1}{n+2} k^2 
	   [ 2 (p \cdot q) (q \cdot k) + q^2 (p \cdot k) ]   }{
          [(p+k)^2+m_{1}^{2}]^2 \,
          (q^2    +m_{2}^{2}) \,
          (r^2    +m_{3}^{2})         }   
  \nonumber \\
   {\cal P}^{0 \, 3}_{2 \, 1 \, 1} \; \rightarrow {\cal H}_{10} 
   & = &
    \int d^{n}p\,d^{n}q\, 
    \frac{ (q \cdot k)^3 
         - \frac{3}{n+2}  k^2 q^2 (q \cdot k)  }{
          [(p+k)^2+m_{1}^{2}]^2 \,
          (q^2    +m_{2}^{2}) \,
          (r^2    +m_{3}^{2})         }   
\end{eqnarray}
For compactness, we omitted in the above formulae the 
mass and momentum arguments of the functions 
${\cal P}^{a \, b}_{2 \, 1 \, 1}(m_1,m_2,m_3;k^2)$ 
and ${\cal H}_i (m_1,m_2,m_3;k^2)$.

We choose to consider this set of scalar integrals because their
ultraviolet behaviour is logarithmic. This is necessary for deriving 
simple integral representations of their ultraviolet finite parts.
It is straightforward to express the ${\cal P}^{a \, b}_{2 \, 1 \, 1}$
functions in terms
of the ${\cal H}_i$ functions. The necessary conversion formulae are
listed in Appendix B.

Simple, one-dimensional representations for the functions 
${\cal H}_1$ and ${\cal H}_2$ were found in ref. \cite{2loop:2}.
As it turns out, it is possible to derive similar integral 
representations for the other functions, ${\cal H}_3$---${\cal H}_{10}$,
as well. We discuss some technical details of deriving such integral
representations in Appendix C. 

By using the methods discussed in Appendix C, one finds the following 
expressions for the functions ${\cal H}_i$ ($\epsilon = n - 4$):

\begin{eqnarray}
   {\cal H}_1 
   & = &
    \pi^4 
    \left[
      \frac{2}{\epsilon^2}
    - \frac{1}{\epsilon} ( 1 - 2 \gamma_{m_1} )
          - \frac{1}{2}
          + \frac{\pi^2}{12}
          - \gamma_{m_1}
          + \gamma_{m_1}^2
          + h_1
   \right]
  \nonumber \\
   {\cal H}_2 
   & = &
    \pi^4 k^2
    \left[
    - \frac{2}{\epsilon^2}
    + \frac{1}{\epsilon} ( \frac{1}{2} - 2 \gamma_{m_1} )
          + \frac{13}{8}
          - \frac{\pi^2}{12}
          + \frac{\gamma_{m_1}}{2}
          - \gamma_{m_1}^2
          - h_2
   \right]
  \nonumber \\
   {\cal H}_3 
   & = &
    \pi^4 k^2
    \left[
      \frac{1}{\epsilon^2}
    - \frac{1}{\epsilon} ( \frac{1}{4} - \gamma_{m_1} )
          - \frac{13}{16}
          + \frac{\pi^2}{24}
          - \frac{\gamma_{m_1}}{4} 
          + \frac{\gamma_{m_1}^2}{2} 
          + h_3
   \right]
  \nonumber \\
   {\cal H}_4 
   & = &
    \pi^4 (k^2)^2
    \left[
      \frac{3}{2 \epsilon^2}
    + \frac{1}{\epsilon} \frac{3 \gamma_{m_1}}{2} 
          - \frac{175}{96}
          + \frac{\pi^2}{16}
          + \frac{3 \gamma_{m_1}^2}{4} 
          + \frac{3}{4} h_4
   \right]
  \nonumber \\
   {\cal H}_5 
   & = &
    \pi^4 (k^2)^2
    \left[
    - \frac{3}{4 \epsilon^2}
    - \frac{1}{\epsilon} \frac{3 \gamma_{m_1}}{4} 
          + \frac{175}{192}
          - \frac{\pi^2}{32}
          - \frac{3 \gamma_{m_1}^2}{8} 
          - \frac{3}{4} h_5
   \right]
  \nonumber \\
   {\cal H}_6 
   & = &
    \pi^4 (k^2)^2
    \left[
      \frac{1}{2 \epsilon^2}
    - \frac{1}{\epsilon} ( \frac{1}{24} - \frac{\gamma_{m_1}}{2}  )
          - \frac{19}{32}
          + \frac{\pi^2}{48}
          - \frac{\gamma_{m_1}}{24} 
          + \frac{\gamma_{m_1}^2}{4} 
          + \frac{3}{4} h_6
   \right]
  \nonumber \\
   {\cal H}_7
   & = &
    \pi^4 (k^2)^3
    \left[
    - \frac{1}{\epsilon^2}
    - \frac{1}{\epsilon} ( \frac{5}{24} + \gamma_{m_1} )
          + \frac{287}{192}
          - \frac{\pi^2}{24}
          - \frac{5 \gamma_{m_1}}{24} 
          - \frac{\gamma_{m_1}^2}{2} 
          - \frac{1}{2} h_7
   \right]
  \nonumber \\
   {\cal H}_8
   & = &
    \pi^4 (k^2)^3
    \left[
      \frac{1}{2\epsilon^2}
    + \frac{1}{\epsilon} ( \frac{5}{48} + \frac{\gamma_{m_1}}{2}  )
          - \frac{287}{384}
          + \frac{\pi^2}{48}
          + \frac{5 \gamma_{m_1}}{48} 
          + \frac{\gamma_{m_1}^2}{4} 
          + \frac{1}{2} h_8
   \right]
  \nonumber \\
   {\cal H}_9 
   & = &
    \pi^4 (k^2)^3
    \left[
    - \frac{1}{3 \epsilon^2}
    - \frac{1}{\epsilon} ( \frac{1}{24} + \frac{\gamma_{m_1}}{3}  )
          + \frac{95}{192}
          - \frac{\pi^2}{72}
          - \frac{\gamma_{m_1}}{24} 
          - \frac{\gamma_{m_1}^2}{6} 
          - \frac{1}{2} h_9
   \right]
  \nonumber \\
   {\cal H}_{10} 
   & = &
    \pi^4 (k^2)^3
    \left[
      \frac{1}{4 \epsilon^2}
    + \frac{1}{\epsilon} ( \frac{1}{96} + \frac{\gamma_{m_1}}{4}  )
          - \frac{283}{768}
          + \frac{\pi^2}{96}
          + \frac{\gamma_{m_1}}{96} 
          + \frac{\gamma_{m_1}^2}{8} 
          + \frac{1}{2} h_{10}
   \right]  
   \; \; .
\end{eqnarray}

Again, to simplify the notations, we omitted in the above formulae
the mass and momentum arguments of the functions 
${\cal H}_i (m_1,m_2,m_3;k^2)$ and $h_i (m_1,m_2,m_3;k^2)$.
$\gamma_{m_1} = \gamma + \log{(\pi m_1^2/\mu_1^2)}$ ,
where $\gamma = 0.577216$ is the Euler constant and $\mu_1$ is 
the 't Hooft mass. The ultraviolet finite 
parts $h_i (m_1,m_2,m_3;k^2)$ of the functions 
${\cal H}_i (m_1,m_2,m_3;k^2)$ have the following one--dimensional 
integral representations:

\begin{eqnarray}
    h_1(m_1,m_2,m_3;k^2) & = &  \int_0^1 dx \,
                                \tilde{g} (x)
  \nonumber \\
    h_2(m_1,m_2,m_3;k^2) & = &  \int_0^1 dx \,
                              [ \tilde{g}   (x)
                              + \tilde{f_1} (x) ]
  \nonumber \\
    h_3(m_1,m_2,m_3;k^2) & = &  \int_0^1 dx \, 
                              [ \tilde{g}   (x)
                              + \tilde{f_1} (x) ] \, (1-x)
  \nonumber \\
    h_4(m_1,m_2,m_3;k^2) & = &  \int_0^1 dx \,
                              [ \tilde{g}   (x)
                              + \tilde{f_1} (x)
                              + \tilde{f_2} (x) ]
  \nonumber \\
    h_5(m_1,m_2,m_3;k^2) & = &  \int_0^1 dx \,
                              [ \tilde{g}   (x)
                              + \tilde{f_1} (x)
                              + \tilde{f_2} (x) ] \, (1-x)
  \nonumber \\
    h_6(m_1,m_2,m_3;k^2) & = &  \int_0^1 dx \,
                              [ \tilde{g}   (x)
                              + \tilde{f_1} (x)
                              + \tilde{f_2} (x) ] \, (1-x)^2
  \nonumber \\
    h_7(m_1,m_2,m_3;k^2) & = &  \int_0^1 dx \,
                              [ \tilde{g}   (x)
                              + \tilde{f_1} (x)
                              + \tilde{f_2} (x)
                              + \tilde{f_3} (x) ]
  \nonumber \\
    h_8(m_1,m_2,m_3;k^2) & = &  \int_0^1 dx \,
                              [ \tilde{g}   (x)
                              + \tilde{f_1} (x)
                              + \tilde{f_2} (x)
                              + \tilde{f_3} (x) ] \, (1-x)
  \nonumber \\
    h_9(m_1,m_2,m_3;k^2) & = &  \int_0^1 dx \,
                              [ \tilde{g}   (x)
                              + \tilde{f_1} (x)
                              + \tilde{f_2} (x)
                              + \tilde{f_3} (x) ] \, (1-x)^2
  \nonumber \\
    h_{10}(m_1,m_2,m_3;k^2) & = &  \int_0^1 dx \,
                              [ \tilde{g}   (x)
                              + \tilde{f_1} (x)
                              + \tilde{f_2} (x)
                              + \tilde{f_3} (x) ] \, (1-x)^3
   \; \; .
\end{eqnarray}

All ten integral representations are built out of the following four
basic functions:

\begin{eqnarray}
  \tilde{g} (m_1,m_2,m_3;k^2;x) & = &
     Sp(\frac{1}{1-y_1}) 
   + Sp(\frac{1}{1-y_2}) 
   + y_1 \log{\frac{y_1}{y_1-1}} 
   + y_2 \log{\frac{y_2}{y_2-1}} 
  \nonumber \\
  \tilde{f_1}(m_1,m_2,m_3;k^2;x) & = &
   \frac{1}{2}
   \left[
   - \frac{1-\mu^2}{\kappa^2}
   + y_1^2 \log{\frac{y_1}{y_1-1}} 
   + y_2^2 \log{\frac{y_2}{y_2-1}} 
   \right]
  \nonumber \\
  \tilde{f_2}(m_1,m_2,m_3;k^2;x) & = &
   \frac{1}{3}
   \left[
   - \frac{2}{\kappa^2} 
   - \frac{1-\mu^2}{2 \kappa^2}
   - \left( \frac{1-\mu^2}{\kappa^2} \right)^2
   \right.
  \nonumber \\
 & &
   \; \; \; \; \; \; \; \; \; \; \; \;
   \; \; \; \; \; \; \; \; \; \; \; \;
   \; \; \; \; \; \; \; \; \; \; \; \;
   \left.
   + y_1^3 \log{\frac{y_1}{y_1-1}} 
   + y_2^3 \log{\frac{y_2}{y_2-1}} 
   \right]
  \nonumber \\
  \tilde{f_3}(m_1,m_2,m_3;k^2;x) & = &
   \frac{1}{4}
   \left[
   - \frac{4}{\kappa^2} 
   - \left( \frac{1}{3} + \frac{3}{\kappa^2}  \right) 
     \left( \frac{1-\mu^2}{\kappa^2} \right)
   - \frac{1}{2} \left( \frac{1-\mu^2}{\kappa^2} \right)^2
   - \left( \frac{1-\mu^2}{\kappa^2} \right)^3
   \right.
  \nonumber \\
 & &
   \left.
   \; \; \; \; \; \; \; \; \; \; \; \;
   \; \; \; \; \; \; \; \; \; \; \; \;
   \; \; \; \; \; \; \; \; \; \; \; \;
   + y_1^4 \log{\frac{y_1}{y_1-1}} 
   + y_2^4 \log{\frac{y_2}{y_2-1}} 
   \right]
   \; \; ,
\end{eqnarray}
where we use the following notations:

\begin{eqnarray}
y_{1,2} & = & \frac{1 + \kappa^{2} - \mu^{2}
                    \pm \sqrt{\Delta}}{2 \kappa^{2}}  \nonumber \\
\Delta  & = & (1 + \kappa^{2} - \mu^{2})^{2} 
          + 4 \kappa^{2} \mu^{2} - 4 i \kappa^{2} \eta 
      \; \; ,
\end{eqnarray}
and

\begin{eqnarray}
   \mu^{2}  & = &  \frac{a x + b (1-x)}{x (1-x)}   \nonumber \\
         a  & = &  \frac{m_{2}^{2}}{m_{1}^{2}} \, , \; \; \; \;
         b \; = \; \frac{m_{3}^{2}}{m_{1}^{2}} \, , \; \; \; \;
\kappa^{2} \; = \; \frac{    k^{2}}{m_{1}^{2}} 
      \; \; .
\end{eqnarray}

In the above expressions, one special case must be treated separately,
namely $k^2 = 0$. One can convince oneself that in this case 
our approach reduces to the functions introduced by van der Bij 
and Veltman in ref. \cite{vdbij}.

The functions $\tilde{g}$ and $\tilde{f_1}$ were already introduced 
in ref. \cite{2loop:2}. They are sufficient for treating Feynman diagrams
without numerators. The other two functions, $\tilde{f_2}$ and 
$\tilde{f_3}$, have a similar structure. They
are needed for treating functions of degree two and one, respectively, 
which may appear for instance in a renormalizable theory with fermions.
Clearly, it is possible to extend these formulae to functions of lower
degree which may appear in nonrenormalizable theories.

In order to calculate more complicated diagrams, the derivatives
of the functions defined in eqns. 20 are needed. One may ask oneself
whether this procedure does not introduce additional singularities in
the integrands, for instance at $x=0,1$. 
   One can convince oneself rather easily that this is not the case. 
   This can be seen by noticing that
   near x = 0 (the situation is the same at x = 1), the function e.g. $\tilde{g}$, 
   behaves like $\log{x}^2$, which is integrable.
   It is easily seen that by differentiating with respect to one mass or
   external momentum the behaviour near $x=0$ will remain integrable.
   By differentiating the function an arbitrary number of times one
   cannot induce additional singularities in the integral representation.
   The derivatives will remain integrable. 
     Since there can be no fundamental difficulty related to the 
   differentiation procedure, the problem of evaluating numerically
   the derivatives is merely a question of implementing the formulae
   in a computer program in a correct, numerically stable form. This
   was shown in previous works, e.g. refs. \cite{2loop:2,2loop:3}, 
   which involve a subset of the $h_i$ functions, and where
   the differentiation procedure was used.

The relations 19 can be used for an efficient numerical evaluation
of two--loop Feynman diagrams. We do not enter here into details 
regarding the numerical integration. Let us just note that this 
was done in refs. \cite{2loop:2}---\cite{2loop:3} for $h_1$ and $h_2$ 
in the case of two-- and three--point functions, and the results
agree with independent calculations \cite{maher}---\cite{frink}. 
The other $h$ functions have analytical structures similar to 
$h_1$ and $h_2$, so that the same numerical algorithms can be used
to evaluate them.


\section{Conclusions}

We presented a method for treating all two--loop diagrams
which may appear in a renormalizable theory in a systematic way. 
Formulae were given which allow one to express any two--loop diagram 
in terms of a small number of scalar functions. We derived 
one--dimensional representations of these scalar functions. 
They are constructed out of four basic functions which have 
quite simple expressions. These integral representations can 
be used further for an efficient evaluation of Feynman diagrams.

A useful feature of our approach is that it is suitable for the 
automatization of two--loop computations. All algebraic 
manipulations needed to reduce a Feynman diagram to scalar 
invariants and to express it in terms of the four basic functions 
which we introduced are algorithmical. They can be encoded in 
a computer algebra program to treat automatically 
two--loop Feynman diagrams.

The calculation of a Feynman diagram implies a numerical integration 
over $h_i$ functions or their derivatives. The maximum dimension of
this final integration is the number of propagators minus three.
Where derivatives of $h_i$ functions are needed, they are 
straightforward to obtain, for instance by using a 
computer algebra program.

The numerical integrations, the details of which 
we do not discuss in this paper, are standard.  
Let us just note that the analytical structure of the functions 
involved in the integral representations is known \cite{2loop:2}. 
This allows one to identify their singularities, and to define a 
complex integration path by means of spline functions. Along such 
a path one can use an adaptative deterministic algorithm for 
a fast and accurate numerical integration \cite{2loop:2,2loop:3}.

In this paper we provide only the formulae which are 
encountered in renormalizable theories. If calculations 
in nonrenormalizable theories are addressed, new structures 
may appear. However, they can be dealt with along the same lines, 
and lead to similar functions. The main difference is that in 
nonrenormalizable theories it may be impossible to isolate a 
finite number of two--loop functions from which all other can be 
derived by differentiation. In this case one would need to 
consider a specific process and to identify the necessary functions.


\vspace{0.5cm}

{\bf Acknowledgement}

We would like to thank the High Energy Theory group of 
Brookhaven National Laboratory for hospitality in the summer 
of 1996. The work of A.G. is supported by the 
Deutsche Forschungsgemeinschaft (DFG).
Y.P.Y.'s work is partially supported by the U.S. 
Department of Energy. 


\appendix
\section*{Appendix A}

We give here the formulae for decomposing a tensor integral 
$I^{\mu_1 \cdots \mu_{j^{\prime}}}$ of the type in eq. 5 for the
case $j^{\prime} = 4$ into scalar invariants.

Apart from a redefinition of the loop momenta $p$ and $q$,
there are three possible cases if  $j^{\prime} = 4$:

\begin{eqnarray}
\lefteqn{\int d^{n}p\,d^{n}q\, 
       \frac{p^{\mu_1}_{\perp} p^{\mu_2}_{\perp} 
             p^{\mu_3}_{\perp} p^{\mu_4}_{\perp}  P(p \cdot k,q \cdot k)}{
             (p^{2}+m_{1}^{2})^{\alpha_{1}} \,
             (q^{2}+m_{2}^{2})^{\alpha_{2}} \,
             [(r+k)^{2}+m_{3}^{2}]^{\alpha_{3}}
	    }
    \, =}  \nonumber \\ 
& &
  C \, \left\{ \left( - k^2 g^{\mu_1 \mu_2} g^{\mu_3 \mu_4} 
                 +     g^{\mu_1 \mu_2} k^{\mu_3} k^{\mu_4}
                 +     g^{\mu_3 \mu_4} k^{\mu_1} k^{\mu_2}
                 - \frac{1}{k^2} k^{\mu_1} k^{\mu_2} k^{\mu_3} k^{\mu_4} 
	  \right) \right.
\nonumber \\ 
& & \left.
  + \left[ ( \mu_1 \mu_2 \mu_3 \mu_4 ) \leftrightarrow 
                              ( \mu_1 \mu_3 \mu_2 \mu_4 ) \right]
  + \left[ ( \mu_1 \mu_2 \mu_3 \mu_4 ) \leftrightarrow 
                              ( \mu_1 \mu_4 \mu_2 \mu_3 ) \right]
   \, \right\}  
\end{eqnarray}

\begin{eqnarray}
\lefteqn{\int d^{n}p\,d^{n}q\, 
       \frac{q^{\mu_1}_{\perp} p^{\mu_2}_{\perp} 
             p^{\mu_3}_{\perp} p^{\mu_4}_{\perp}  P(p \cdot k,q \cdot k)}{
             (p^{2}+m_{1}^{2})^{\alpha_{1}} \,
             (q^{2}+m_{2}^{2})^{\alpha_{2}} \,
             [(r+k)^{2}+m_{3}^{2}]^{\alpha_{3}}
	    }
    \, =}  \nonumber \\ 
& &
 D \, \left\{ \left( - k^2 g^{\mu_1 \mu_2} g^{\mu_3 \mu_4} 
                 +     g^{\mu_1 \mu_2} k^{\mu_3} k^{\mu_4}
                 +     g^{\mu_3 \mu_4} k^{\mu_1} k^{\mu_2}
                 - \frac{1}{k^2} k^{\mu_1} k^{\mu_2} k^{\mu_3} k^{\mu_4} 
	  \right) \right.
\nonumber \\ 
& & \left.
  + \left[ ( \mu_1 \mu_2 \mu_3 \mu_4 ) \leftrightarrow 
                              ( \mu_1 \mu_3 \mu_2 \mu_4 ) \right]
  + \left[ ( \mu_1 \mu_2 \mu_3 \mu_4 ) \leftrightarrow 
                              ( \mu_1 \mu_4 \mu_2 \mu_3 ) \right]
   \, \right\} 
\end{eqnarray}

\begin{eqnarray}
\lefteqn{\int d^{n}p\,d^{n}q\, 
       \frac{q^{\mu_1}_{\perp} q^{\mu_2}_{\perp} 
             p^{\mu_3}_{\perp} p^{\mu_4}_{\perp} P(p \cdot k,q \cdot k)}{
             (p^{2}+m_{1}^{2})^{\alpha_{1}} \,
             (q^{2}+m_{2}^{2})^{\alpha_{2}} \,
             [(r+k)^{2}+m_{3}^{2}]^{\alpha_{3}}
	    }
    \, =}  \nonumber \\ 
& &
   \left\{ E \, \left( - k^2 g^{\mu_1 \mu_2} g^{\mu_3 \mu_4} 
                 +     g^{\mu_1 \mu_2} k^{\mu_3} k^{\mu_4}
                 +     g^{\mu_3 \mu_4} k^{\mu_1} k^{\mu_2}
                 - \frac{1}{k^2} k^{\mu_1} k^{\mu_2} k^{\mu_3} k^{\mu_4} 
	  \right) \right. 
\nonumber \\ 
& & + \left. F \, \left\{
    \left[ ( \mu_1 \mu_2 \mu_3 \mu_4 ) \leftrightarrow 
                              ( \mu_1 \mu_3 \mu_2 \mu_4 ) \right] 
  + \left[ ( \mu_1 \mu_2 \mu_3 \mu_4 ) \leftrightarrow 
                              ( \mu_1 \mu_4 \mu_2 \mu_3 ) \right] 
   \, \right\} \right\} 
    \; \; ,
\end{eqnarray}
where

\begin{eqnarray}
C & = &  - \frac{1}{(n^2-1) k^2}
\int d^{n}p\,d^{n}q\, 
       \frac{  (p_{\perp}^2)^2 
             P(p \cdot k , q \cdot k)}{
             (p^{2}+m_{1}^{2})^{\alpha_{1}} \,
             (q^{2}+m_{2}^{2})^{\alpha_{2}} \,
             [(r+k)^{2}+m_{3}^{2}]^{\alpha_{3}}}
\nonumber \\ 
D & = &  - \frac{1}{(n^2-1) k^2}
\int d^{n}p\,d^{n}q\, 
       \frac{ (p_{\perp} \cdot q_{\perp})
	       p_{\perp}^2
             P(p \cdot k , q \cdot k)}{
             (p^{2}+m_{1}^{2})^{\alpha_{1}} \,
             (q^{2}+m_{2}^{2})^{\alpha_{2}} \,
             [(r+k)^{2}+m_{3}^{2}]^{\alpha_{3}}}
\nonumber \\ 
E & = &  - \frac{1}{(n^2-1) (n-2) k^2}\times
\nonumber \\ 
 & & \; \;\; \;\; \;\; \;\; \;\; \;\; \;\; \;\; \;\; \;
       \int d^{n}p\,d^{n}q\, 
       \frac{ \left[ 
              n p_{\perp}^2 q_{\perp}^2 - 2 (p_{\perp} \cdot q_{\perp})^2
              \right] 
             P(p \cdot k , q \cdot k)}{
             (p^{2}+m_{1}^{2})^{\alpha_{1}} \,
             (q^{2}+m_{2}^{2})^{\alpha_{2}} \,
             [(r+k)^{2}+m_{3}^{2}]^{\alpha_{3}}}
\nonumber \\ 
F& = &    \frac{1}{(n^2-1) (n-2) k^2} \times
\nonumber \\ 
 & & \; \;\; \;\; \;\; \;\; \;\; \;\; \;\; \;\; \;\; \;
       \int d^{n}p\,d^{n}q\, 
       \frac{ \left[ 
              p_{\perp}^2 q_{\perp}^2 - (n-1) (p_{\perp} \cdot q_{\perp})^2
              \right] 
             P(p \cdot k , q \cdot k)}{
             (p^{2}+m_{1}^{2})^{\alpha_{1}} \,
             (q^{2}+m_{2}^{2})^{\alpha_{2}} \,
             [(r+k)^{2}+m_{3}^{2}]^{\alpha_{3}}}
    \; \; . 
\end{eqnarray}

The scalar invariants $C$, $D$, $E$ and $F$ can be further reduced 
to functions of the form of eq. 3 by using the relations:

\begin{eqnarray}
p_{\perp}^2               & = & p^2       - \frac{1}{k^2} \, 
                                                 (p \cdot k)^2
\nonumber \\ 
p_{\perp} \cdot q_{\perp} & = & p \cdot q - \frac{1}{k^2} \,
                                      (p \cdot k)(q \cdot k)
\nonumber \\ 
p \cdot q                 & = & \frac{1}{2} \, (r^2 - p^2 - q^2)
\nonumber 
\end{eqnarray}
and by partial fractioning the resulting expressions.


\appendix
\section*{Appendix B}

In section 4 we replaced the set of functions 
${\cal P}^{a \, b}_{2 \, 1 \, 1}$ , $a+b = 0, 1, 2, 3$
by the functions ${\cal H}_1$---${\cal H}_{10}$.
The functions ${\cal H}_i$ are free of quadratic
divergencies, and for this reason they have simpler
integral representations. The relations between 
${\cal P}^{a \, b}_{2 \, 1 \, 1}$ and ${\cal H}_i$
are easily obtained by partial fractioning the eqns. 17:

\begin{eqnarray}
    {\cal P}^{0 \, 0}_{2 \, 1 \, 1} & = &
    {\cal H}_1
  \nonumber \\
    {\cal P}^{1 \, 0}_{2 \, 1 \, 1} & = &
    - {\cal H}_2 - k^2 {\cal H}_1
  \nonumber \\
    {\cal P}^{0 \, 1}_{2 \, 1 \, 1} & = & -
    {\cal H}_3
  \nonumber \\
    {\cal P}^{2 \, 0}_{2 \, 1 \, 1} & = &
    {\cal H}_4 + \frac{k^2}{n} \left\{
    [ (n-1) k^2 - m_1^2 ] {\cal H}_1
  + 2 (n-1) {\cal H}_2
  + {\cal P}^{0 \, 0}_{1 \, 1 \, 1}
                              \right\}
  \nonumber \\
    {\cal P}^{1 \, 1}_{2 \, 1 \, 1} & = &
      {\cal H}_5 
    + k^2 {\cal H}_3
    + \frac{k^2}{2 n} \left\{
    (m_1^2 + m_2^2 - m_3^2 + k^2) {\cal H}_1
  + 2 {\cal H}_2
  - {\cal P}^{0 \, 0}_{1 \, 1 \, 1}   \right.
  \nonumber \\
   &   &   \left.
  \; \; \; \; \; \; \; \; \; \; 
  \; \; \; \; \; \; \; \; \; \; 
  \; \; \; \; \; \; \; \; \; \; 
  \; \; \; \; \; \; \; \; \; \; 
  \; \; \; \; \; \; \; \; \; \; 
  \; \; \; \; \; \; \; \; \; \; 
  \; \; \; \; \; \; \; \; 
  + T_2(m_1^2) \left[ T_1(m_2^2) - T_1(m_3^2) \right]
                              \right\}
  \nonumber \\ 
    {\cal P}^{0 \, 2}_{2 \, 1 \, 1} & = &
    {\cal H}_6 + \frac{k^2}{n} \left[
  - m_2^2  {\cal H}_1
  + T_2(m_1^2) T_1(m_3^2) 
                              \right]
  \nonumber \\
    {\cal P}^{3 \, 0}_{2 \, 1 \, 1} & = &
  - {\cal H}_7 
  - \frac{3 k^2}{n+2} \left\{
    \left( \frac{n-1}{3} k^2 - m_1^2 \right) k^2 {\cal H}_1
  + [(n-1) k^2 - m_1^2 ] {\cal H}_2
  + n {\cal H}_4
  - {\cal P}^{1 \, 0}_{1 \, 1 \, 1}
                              \right\}
  \nonumber \\
    {\cal P}^{2 \, 1}_{2 \, 1 \, 1} & = &
  - {\cal H}_8 
  - \frac{3 k^2}{n+2} \left[
     \frac{2}{3} (n-1) {\cal H}_5
  +  \left(\frac{n-1}{3} k^2 - m_1^2 \right)  {\cal H}_3  
  -  {\cal P}^{0 \, 1}_{1 \, 1 \, 1}
                      \right]
  - \frac{n-1}{n (n+2)} (k^2)^2  
  \nonumber \\
   &   &   
  \times \left\{
    (m_1^2 + m_2^2 - m_3^2 + k^2) {\cal H}_1
  + 2 {\cal H}_2
  - {\cal P}^{0 \, 0}_{1 \, 1 \, 1}
  + T_2(m_1^2) [T_1(m_2^2) - T_1(m_3^2)]
                              \right\}
  \nonumber \\
    {\cal P}^{1 \, 2}_{2 \, 1 \, 1} & = &
  - {\cal H}_9 
  - k^2 {\cal H}_6  
  - \frac{k^2}{n+2} \left[
    2 {\cal H}_5
  + (\frac{2k^2}{n} - m_2^2){\cal H}_2
  + (m_1^2 + m_2^2 - m_3^2 + k^2) {\cal H}_3
  + {\cal P}^{0 \, 1}_{1 \, 1 \, 1} 
                              \right]
  \nonumber \\
   &   &   
  - \frac{(k^2)^2}{n (n+2)} \left\{
    (m_1^2 - (n+1) m_2^2 - m_3^2 + k^2){\cal H}_1
  - {\cal P}^{0 \, 0}_{1 \, 1 \, 1}  \right.
  \nonumber \\
   &   &   \left.
  \; \; \; \; \; \; \; \; \; \; 
  \; \; \; \; \; \; \; \; \; \; 
  \; \; \; \; \; \; \; \; \; \; 
  \; \; \; \; \; \; \; \; \; \; 
  \; \; \; \; \; \; \; \; \; \; 
  \; \; \; \; \; \; \; \; 
  + T_2(m_1^2) [ T_1(m_2^2) - (n-1) T_1(m_3^2) ]
                              \right\}
  \nonumber \\
    {\cal P}^{0 \, 3}_{2 \, 1 \, 1} & = &
  - {\cal H}_{10} - \frac{3 k^2}{n+2} \left[
  - m_2^2 {\cal H}_3
  + k^2 T_2(m_1^2) T_1(m_3^2)     
                              \right]
  \; \; \; \; ,
\end{eqnarray}
where $T_1$ and $T_2$ are the Euclidian one--loop tadpole integrals:

\begin{eqnarray}
   T_1(m^2) & = & \int d^{n}p\, \frac{1}{p^2+m^2}  
    \; = \; 
   - \pi^2 
     \left( \pi m^2 \right)^{\frac{\epsilon}{2}}
     \Gamma\left( - \frac{\epsilon}{2} \right) \,
     \frac{2 m^2}{2+\epsilon} 
  \nonumber \\
   T_2(m^2) & = & \int d^{n}p\, \frac{1}{(p^2+m^2)^2}  
    \; = \; 
     \pi^2 
     \left( \pi m^2 \right)^{\frac{\epsilon}{2}}
     \Gamma\left( - \frac{\epsilon}{2} \right)
    \; \; .
\end{eqnarray}

For simplifying the notation, we omitted in the above formulae
the mass and momentum arguments of the functions
${\cal H}_i(m_1,m_2,m_3;k^2)$ and
${\cal P}^{a \, b}_{\alpha_1\, \alpha_2 \, \alpha_3}
(m_1,m_2,m_3;k^2)$, 
and understand that these arguments appear in 
this order in all relations.

The functions with $\alpha_1=\alpha_2=\alpha_3=1$ which appear in
the relations 27 can be calculated by partial $p$ (eq. 15):

\begin{eqnarray}
  {\cal P}^{1 \, 0}_{1 \, 1 \, 1} (m_1,m_2,m_3;k^2) & = &
  - \frac{1}{n-\frac{5}{2}} \left\{
    \left[
         \frac{k^2}{2}
	 {\cal P}^{0 \, 0}_{1 \, 1 \, 1}
       - {\cal P}^{2 \, 0}_{2 \, 1 \, 1}
       - m_1^2 ({\cal H}_2 + k^2 {\cal H}_1)
    \right] (m_1,m_2,m_3;k^2)
  \right.
  \nonumber \\
   &   &   
   \left.
                    - m_2^2 {\cal H}_3  (m_2,m_1,m_3;k^2)
      + m_3^2 ({\cal H}_2 + {\cal H}_3) (m_3,m_2,m_1;k^2)
	                  \right\}
 \nonumber \\
  {\cal P}^{0 \, 0}_{1 \, 1 \, 1} (m_1,m_2,m_3;k^2) & = &
  - \frac{1}{n-3} \left\{
  \left[ 
        (m_1^2 + k^2) {\cal H}_1 + {\cal H}_2
  \right] (m_1,m_2,m_3;k^2)
  \right.
  \nonumber \\
   &   &   
  \; \; \; \; \;\; \; \; \; \;
   \left.
        + m_2^2 {\cal H}_1 (m_2,m_1,m_3;k^2)
        + m_3^2 {\cal H}_1 (m_3,m_1,m_2;k^2)
                   \right\}
\end{eqnarray}


\appendix
\section*{Appendix C}

Here we make some comments on the derivation of the integral 
representations of the functions ${\cal H}_i$. The derivation
proceeds in the same way for all ten functions, and works
for functions of lower degree as well. 

Let us consider for instance the function ${\cal H}_6$:

\begin{equation}
   {\cal H}_6 (m_1,m_2,m_3;k^2) = 
    \int d^{n}p\,d^{n}q\, 
    \frac{ (q \cdot k)^2 
         - \frac{1}{n} k^2 q^2 }{
          [(p+k)^2+m_{1}^{2}]^2 \,
          (q^2    +m_{2}^{2}) \,
          (r^2    +m_{3}^{2})         }   
\end{equation}

We introduce two Feynman parameters $x$ and $y$, and integrate out
the loop momenta $p$ and $q$. Note that the quadratic divergence
which is present in the ${\cal P}^{0 \, 2}_{2 \, 1 \, 1}$ integral
cancels out in ${\cal H}_6$. This is necessary for obtaining a 
well--defined integral representation of the ultraviolet finite part.
Without this cancellation, part of the ultraviolet divergence of the
diagram is transferred from the radial loop momenta integration to the
Feynman parameters integration; one is then left with a nonintegrable 
singularity in the $x$ integration. 

After integrating out the $q$ and $p$ loop momenta, 
one finds ($n = 4+\epsilon$):

\begin{eqnarray}
   {\cal H}_6 (m_1,m_2,m_3;k^2) & = & 
   \pi^4 (k^2)^2 \Gamma \left( - \frac{\epsilon}{2} \right)
                 \Gamma ( - \epsilon )
		 \frac{1 - \frac{1}{n}}{\Gamma\left(3-\frac{n}{2}\right)}
                 \left( \pi m_1^2 \right)^{\epsilon}  
  \nonumber \\
  & &
 \times \int_0^1 dx \, [x(1-x)]^{\frac{\epsilon}{2}} (1-x)^2 
  \nonumber \\
  & &
 \times \int_0^1 dy \, \left[ \frac{R^2}{1-y} \right]^{\frac{\epsilon}{2}} 
                       \left[ 4 + \frac{\epsilon}{2}
		         \left( 1 - 2 \frac{L}{R} \right) \right] y^3
  \; \; \; .
\end{eqnarray}

Here we introduced the following notation in addition to eqns. 21 and 22:

\begin{eqnarray}
L & = & y^2 \kappa^2 + \mu^2  \nonumber \\
R & = & y (1-y) \kappa^2 + y + (1-y) \mu^2  
        \; \equiv \; \kappa^2 (y_1-y)(y-y_2)
\end{eqnarray}

Eq. 31 displays already the structure of the function ${\cal H}_6$.
The first line of eq. 31 contains the ultraviolet singularity, while
the $y$ and $x$ integrals are finite. Hence, the $x$ and $y$ 
integrations need to be performed up to ${\cal O}(\epsilon^2)$.
Therefore, one expands the integrand in powers of
$\epsilon$ up to order ${\cal O}(\epsilon^2)$ and carries out the $y$
integration. The calculation can be simplified considerably 
by using the relation:

\begin{equation}
  \frac{L}{R} = 1 - \frac{y}{R} \frac{\partial R}{\partial y}
  \; \; \; .
\end{equation}

Finally, one obtains after the $y$ integration the results 
given in eqns. 18---20.



\end{document}